\documentstyle[psfig,epsfig]{article}
\newcommand{\dd}{\mbox{{\rm d}}}

\newcommand{\Lumint}{{\cal L}_{\rm int}}

\begin{document}

\begin{center}

{\Large \bf Prospects on Compositeness and New Vector Bosons at LC with
Polarization\footnote
{Talk given at the International Workshop QFTHEP'99, Moscow, May 27 - June 2,
1999}
} \\

\vspace{4mm}
A.A. Pankov\\
Technical University of Gomel, 246746 Gomel \\
Belarus \\
\end{center}

\begin{abstract}

Fermion compositness, and other new physics which can be described by an 
exchange of very massive particles ($Z^\prime$ boson, leptoquarks,  
sparticles with $R$-parity violating couplings), can be manifest itself 
as the presence of a strong four-fermion contact interaction.
For the processes $e^+e^-\to \mu^+\mu^-$, $\tau^+\tau^-$, $\bar{b}b$
and $\bar{c}c$ at a future $e^+e^-$ linear
collider (LC) with $\sqrt{s}=0.5$ TeV, we examine the sensitivity of the 
helicity cross sections to four-fermion contact interactions. If longitudinal
polarization of the electron beam were available, two polarized integrated
cross sections would offer the opportunity to separate the helicity cross
sections and, in this way, to derive model-independent bounds on the relevant
parameters. The measurement of these polarized cross sections with optimal
kinematical cuts could significantly increase the sensitivity of helicity cross
sections to contact interaction parameters and could give crucial 
information on the chiral structure of such new interactions.
In addition, we consider the application of the proposed approach to the search 
for manifestations of a $Z^\prime$ for typical extended model examples.
\end{abstract}


\section{Introduction and separation of the helicity cross sections}

Deviations from the Standard Model (SM) caused by new physics characterized 
by very high mass scales $\Lambda$ can systematically be studied at lower
energies by using the effective Lagrangian approach. 
The effects of the new physics can be observed at energies well-below 
$\Lambda$, and can be related to some effective contact 
interaction \cite{Eichten, Ruckl}. 
In the framework of composite models of leptons and quarks, the contact 
interaction is regarded as a remnant of the binding force between the 
fermion substructure constituents. Furthermore, in $e^+e^-$ collisions, 
many types of new physics, for which the exchanged particles in the 
$s$, $t$, or $u$ 
channels have mass-squared much larger than the corresponding 
Mandelstam invariant variables, can be described by an effective $eeff$ 
contact term in the interaction Lagrangian \cite{Schrempp}-\cite{Zerwas1}.
For example, effects of a $Z^\prime$ boson of a few TeV mass scale would be 
well-represented by a four-fermion contact interaction. The exchange of a 
leptoquark of a similar mass scale could be described by an effective 
$eeqq$ contact term in the relevant interaction. 
At energies much lower than
the sparticle masses, $R$-parity breaking interactions introduce effective 
$eell$ and $eeqq$ interactions. Thus, quite generally, the 
contact interaction is considered as a convenient parameterization of 
deviations from the SM that may be caused by some new physics at the large
scale~$\Lambda$. 

Fermion-pair production in $e^+e^-$ collisions 
\begin{equation}
e^++e^-\to \bar{f}+f \label{proc}
\end{equation}
($f=l$ or $q$) is one of the basic processes of the SM, and deviations of
the measured observables from the predicted values would be a first 
indication of new physics beyond the SM.

The lowest-order four-fermion contact terms have dimension $D=6$,
which implies
that they are suppressed by $g^2_{\rm eff}/\Lambda^2$. Restricting the fermion
currents to be helicity conserving and flavor diagonal, the general
$SU(3)\times SU(2)\times U(1)$ invariant four-fermion $eeff$ contact
interaction Lagrangian with $D=6$ can be written as
\cite{Eichten}--\cite{Barger2}:
\begin{eqnarray}
{{\cal L}}=\frac{g^2_{\rm eff}}{\Lambda^2}\left[
\eta_{\rm LL}\left(\bar e_{\rm L}\gamma_\mu e_{\rm L}\right)
\left(\bar f_{\rm L}\gamma^\mu f_{\rm L}\right)
+\eta_{\rm LR}\left(\bar{e}_{\rm L}\gamma_\mu e_{\rm L}\right)
\left(\bar f_{\rm R}\gamma^\mu f_{\rm R}
\right)\right. \nonumber \\
+\left.\eta_{\rm RL}\left(\bar{e}_{\rm R}\gamma_\mu e_{\rm R}\right)
\left(\bar f_{\rm L}
\gamma^\mu f_{\rm L}\right) +
\eta_{\rm RR}\left(\bar e_{\rm R}\gamma_\mu e_{\rm R}\right)
\left(\bar f_{\rm R}\gamma^\mu f_{\rm R}\right)
\right],
\label{lagra}
\end{eqnarray}
where generation and color indices have been suppressed. 
It is conventional to take
$g^2_{\rm eff}=4\pi$ and $\eta_{\alpha\beta}=\pm 1$ ($\alpha,\beta=$ L, R).

In general, for a given fermion flavor $f$, Eq.~(\ref{lagra}) defines eight
independent, individual interaction models corresponding to the combinations
of the four chiralities LL, LR, RL and RR with the $\pm$ signs of the
$\eta$'s. In practice, the true interaction might correspond to one of these
models or to any combination of them. 
Here, we will perform a model-independent analysis of the
contact interactions.

In principle, the sought-for deviations of observables from the SM predictions,
giving information on $\Lambda$'s, simultaneously depend on all four-fermion
effective coupling constants in Eq.~(\ref{lagra}), which therefore cannot be
easily disentangled. For simplicity, the analysis is usually performed by
taking a non-zero value for only one parameter at a time,
all the remaining ones being put equal to zero.
Limits on individual $eeqq$ contact interaction parameters have recently 
been derived by this procedure,
from a global analysis of the relevant 
data \cite{Barger2}, and the individual models are severely constrained, 
with $\Lambda_{\alpha\beta}\sim{\cal{O}}(10)$ TeV.
However, if several terms of different chiralities were simultaneously taken
into account, cancellations may occur and the resulting bounds on
$\Lambda_{\alpha\beta}$ would be considerably weaker, of the order of
$3-4$ TeV. 
Consequently, a definite improvement of the situation in this regard 
should be obtained from a procedure of analyzing experimental data that
allows to account for the various contact interaction couplings
simultaneously as
free parameters, and yet to obtain in a model-independent way separate bounds
for the corresponding $\Lambda$'s, not affected by possible accidental
cancellations.
In this paper we shall propose an analysis of $eell$, $eebb$
and $eecc$ contact interactions at the next linear $e^+e^-$ collider 
with $\sqrt s= 500$ GeV and with longitudinally polarized beams.
Our approach makes use of two particular, polarized, integrated cross 
sections $\sigma_1$ and $\sigma_2$, that
are directly connected, {\it via} linear combinations, to the helicity cross
sections of process (\ref{proc}), and therefore allow to deal with a minimal
set of independent free parameters.

This kind of observables, defined for specific kinematical cuts, were already
introduced to study $Z^\prime$ signals at LEP2 and LC \cite{Osland,Babich} 
and
potential manifestations of compositeness at the LC \cite{Pankov}. Here, we
extend the previous considerations by performing a general analysis where,
in the definition of the above-mentioned integrated observables, we choose
suitable kinematical regions where the sensitivity to individual four-fermion
contact interaction parameters is maximal.
Also, an application of the proposed approach to $Z^\prime$ 
searches at LC is discussed where, in particular, the constraints on 
the relevant parameters are derived.  
 
In the Born approximation, including the $\gamma$ and $Z$ exchanges
as well as the four-fermion
contact interaction term (\ref{lagra}), but neglecting $m_f$ with respect
to the c.m.\ energy $\sqrt s$, the differential cross section for the process
$e^+e^-\to  f\bar{f}$ ($f\neq e, t$)
with longitudinally polarized electron-positron beams,
can be written as 
\begin{equation}
\frac{\dd\sigma}{\dd\cos\theta}
=\frac{3}{8}
\left[(1+\cos\theta)^2 {\sigma}_+
+(1-\cos\theta)^2 {\sigma}_-\right],
\label{cross}
\end{equation}
where $\theta$ is the angle between the incoming electron and the outgoing
fermion in the c.m.\ frame.
The functions $\sigma_\pm$ can be expressed in terms of helicity
cross sections
\begin{equation}
\sigma_{\alpha\beta}=N_C\sigma_{\rm pt}
\vert A_{\alpha\beta}\vert^2.
\label{helcross}
\end{equation}
Here, $N_C$ is the QCD factor: $N_C\approx 3(1+\alpha_s/\pi)$
for quarks and  $N_C=1$ for leptons, respectively, and
$\sigma_{\rm pt}\equiv\sigma(e^+e^-\to\gamma^\ast\to l^+l^-)
=(4\pi\alpha^2)/(3s)$.
With electron and positron longitudinal polarizations $P_e$ and $P_{\bar e}$,
the relations are
\begin{eqnarray}
{\sigma}_{+}&=&\frac{1}{4}\,
\left[(1-P_e)(1+P_{\bar{e}})\,\sigma_{\rm LL}
+(1+P_e)(1- P_{\bar{e}})\,\sigma_{\rm RR}\right], \label{s+} \\
{\sigma}_{-}&=&\frac{1}{4}\,
\left[(1-P_e)(1+ P_{\bar{e}})\,\sigma_{\rm LR}
+(1+P_e)(1-P_{\bar{e}})\,\sigma_{\rm RL}\right]. \label{s-}
\end{eqnarray}
The helicity amplitudes
$A_{\alpha\beta}$ can be written as
\begin{equation}
A_{\alpha\beta}=Q_e Q_f+g_\alpha^e\,g_\beta^f\,\chi_Z+
\frac{s\eta_{\alpha\beta}}{\alpha\Lambda_{\alpha\beta}^2},
\label{amplit}
\end{equation}
where the gauge boson propagator is $\chi_Z=s/(s-M^2_Z+iM_Z\Gamma_Z)$,
the SM left- and right-handed fermion couplings of the $Z$ are
$g_{\rm L}^f=(I_{3L}^f-Q_f s_W^2)/s_W c_W$ and
$g_{\rm R}^f=-Q_f s_W^2/s_W c_W$ with
$s_W^2=1-c_W^2\equiv \sin^2\theta_W$, and $Q_f$ are the fermion electric
charges.

The total cross section and the difference of forward and backward cross
sections are given as
\begin{eqnarray}
\label{canon}
\sigma
&=&{\sigma}_{+}+{\sigma}_{-}
=\frac{1}{4}\left[(1-P_e)(1+P_{\bar{e}})(\sigma_{\rm LL}+\sigma_{\rm LR})
+(1+P_e)(1- P_{\bar{e}})(\sigma_{\rm RR}+\sigma_{\rm RL})\right], \\
\sigma_{\rm FB}
&\equiv&\sigma_{\rm F}-\sigma_{\rm B}
=\frac{3}{4}\left({\sigma}_{+}-{\sigma}_{-}\right) \nonumber \\
&=&\frac{3}{16}\left[(1-P_e)(1+P_{\bar{e}})(\sigma_{\rm LL}-\sigma_{\rm LR})
+(1+P_e)(1- P_{\bar{e}})(\sigma_{\rm RR}-\sigma_{\rm RL})\right],
\end{eqnarray}
where
\begin{equation}
\label{sfb}
\sigma_{\rm F}
=\int_{0}^{1}(\dd\sigma/\dd\cos\theta)\dd\cos\theta, \qquad
\sigma_{\rm B}
=\int_{-1}^{0}(\dd\sigma/\dd\cos\theta)\dd\cos\theta.
\end{equation}
Taking Eq.~(\ref{amplit}) into account, these relations show that in general 
$\sigma$ and $\sigma_{\rm FB}$ simultaneously involve all contact-interactions 
couplings even in the polarized case. Therefore, by themselves, these 
measurements do not allow a completely model-independent analysis avoiding, 
in particular, potential cancellations among different couplings. 

Our analysis will be based on the consideration of the four helicity cross 
sections $\sigma_{\alpha\beta}$ as the basic independent observables to be
measured from data on the polarized differential cross section. These cross 
sections depend each on just one individual four-fermion contact parameter 
and therefore lead to a model-independent analysis where all 
$\eta_{\alpha\beta}$ can be taken simultaneously into account as completely 
free parameters, with no danger from potential cancellations. As 
Eqs.~(\ref{s+}) and (\ref{s-}) show, helicity cross sections can be 
disentangled {\it via} the measurement of ${\sigma}_{+}$ and ${\sigma}_{-}$ 
with different choices of the initial beam polarizations.  

One possibility is to project out ${\sigma}_+$ and ${\sigma}_-$ from 
$\dd\sigma/\dd\cos\theta$, as differences of integrated
cross sections. To this aim, we define $z^*_\pm\equiv\cos\theta^*_\pm$ such
that
\begin{equation}
\label{zpm}
\left(\int_{z^*_\pm}^1-\int_{-1}^{z^*_\pm}\right)
\left(1\mp\cos\theta\right)^2\dd\cos\theta=0.
\end{equation}
One finds the solutions:
$z^*_\pm=\mp(2^{2/3}-1)=\mp 0.587$ ($\theta^*_+=126^\circ$ and
$\theta^*_-=54^\circ$). These values satisfy
$(z_\pm^*\mp1)^3=\mp4$. In the case of
$\vert\cos\theta\vert=c<1$, one has ${\vert z^*_\pm\vert=(1+3c^2)^{1/3}-1}$.

From Eq.~(\ref{cross}) one can easily see that at these values of $z^*_\pm$
the difference of two integrated cross sections defined as
\begin{equation}
\label{proj}
\sigma_1(z^*_\pm)-\sigma_2(z^*_\pm)
\equiv \left(\int_{z^*_\pm}^1-\int_{-1}^{z^*_\pm}\right)
\frac{\dd\sigma}{\dd\cos\theta}\,\dd\cos\theta
\end{equation}
is directly related to $\sigma_\pm$ as:
\begin{equation}
\label{tildesigma+-}
\sigma_1(z^*_+)-\sigma_2(z^*_+)=\gamma\sigma_{+}, \qquad 
\sigma_2(z^*_-)-\sigma_1(z^*_-)=\gamma\sigma_{-},
\end{equation}
where $\gamma=3\left(2^{2/3}-2^{1/3}\right)=0.982$.

The solutions of the system of two equations corresponding to $P_e=\pm P$,
and assuming  unpolarized positrons $P_{\bar e}=0$,
in Eqs.~(\ref{s+}) and (\ref{s-}), can be written as:
\begin{eqnarray}
\label{SLL}
\sigma_{\rm LL}
&=&\frac{1+P}{P}\sigma_{+}(-P)-\frac{1-P}{P}\sigma_{+}(P), \\
\label{SRR}
\sigma_{\rm RR}
&=&\frac{1+P}{P}\sigma_{+}(P)-\frac{1-P}{P}\sigma_{+}(-P), \\
\label{SLR}
\sigma_{\rm LR}
&=&\frac{1+P}{P}\sigma_{-}(-P)-\frac{1-P}{P}\sigma_{-}(P), \\
\label{SRL}
\sigma_{\rm RL}
&=&\frac{1+P}{P}\sigma_{-}(P)-\frac{1-P}{P}\sigma_{-}(-P).
\end{eqnarray}
From Eqs. (\ref{SLL})--(\ref{SRL}) one can easily see that
this procedure allows to extract $\sigma_{\rm LL}$, $\sigma_{\rm RR}$,
$\sigma_{\rm LR}$ and $\sigma_{\rm RL}$ by the four independent
measurements of
$\sigma_1(z^*_\pm)$ and $\sigma_2(z^*_\pm)$ at $P_e=\pm P$.

\section{Optimization and model independent analysis}
This extraction of helicity cross sections can be obtained more generally.
Indeed, let us divide the full angular range,
$\vert\cos\theta\vert\le 1$ into two parts, ($-1,\ z^*$) and
($z^*,\ 1$), with arbitrary $z^*$, and define two integrated cross sections as
\begin{eqnarray}
\label{sigma1}
\sigma_1(z^*)
&\equiv&\int_{z^*}^1\frac{\dd\sigma}{\dd\cos\theta}\dd\cos\theta
=\frac{1}{8}\left\{\left[8-(1+z^*)^3\right]\sigma_++(1-z^*)^3
\sigma_-\right\}, \\
\label{sigma2}
\sigma_2(z^*)
&\equiv&\int^{z^*}_{-1}\frac{\dd\sigma}{\dd\cos\theta}\dd\cos\theta
=\frac{1}{8}\left\{(1+z^*)^3\sigma_++
\left[8-(1-z^*)^3\right]\sigma_-\right\}.
\end{eqnarray}
Solving these two equations, one finds the general relations
\begin{eqnarray}
\label{sigmap}
\sigma_+
&=&\frac{1}{6(1-{z^*}^2)}\left[\left(8-(1-z^*)^3\right)
\sigma_1(z^*)-(1-z^*)^3\sigma_2(z^*)\right], \\
\label{sigmam}
\sigma_-
&=&\frac{1}{6(1-{z^*}^2)}\left[-(1+z^*)^3\sigma_1(z^*)
+\left(8-(1+z^*)^3\right)\sigma_2(z^*)\right],
\end{eqnarray}
that allow to disentangle the
helicity cross sections, using (\ref{SLL})--(\ref{SRL}) and
the availability of polarized beams.

We take radiative corrections into account by means of the program
ZFITTER \cite{zfitter}, which has to be used along with ZEFIT,
adapted to the present discussion.
Due to the radiative return to the $Z$ resonance at $\sqrt{s}>M_Z$, the
energy spectrum of the radiated photons is peaked around
$k_{\rm peak}\approx 1-M^2_Z/s$ \cite{Djouadi}.
In order to increase the signal originating from contact interactions,
events with hard photons should be eliminated
by an appropriate cut $\Delta<k_{\rm peak}$ on the photon energy.
For our numerical analysis, we use $m_{\rm top}=175$~GeV, $m_H=100$~GeV and 
a cut $\sqrt{s^\prime}\ge 0.9\sqrt{s}$ to avoid the radiative return to 
the $Z$ peak for $\sqrt{s}=0.5$ TeV.

In the case where no deviation from the SM is observed, one can make an
assessment of the sensitivity of the process (\ref{proc}) to the contact 
interaction parameters, based on the expected experimental accuracy on the 
observables $\sigma_{\alpha\beta}$. Such sensitivity numerically determines 
the bounds on the contact-interaction scales $\Lambda_{\alpha\beta}$ 
that can be derived from the experimental data and, basically, 
is determined by the comparison
of deviations from the SM predictions due to the contact-interaction terms 
with the attainable experimental uncertainty. Accordingly, we define the 
`significance' of each helicity cross section by the ratio:
\begin{equation}
\label{signif}
{\cal S}
=\frac{|\Delta\sigma_{\alpha\beta}|}{\delta\sigma_{\alpha\beta}}, 
\end{equation}
where $\Delta\sigma_{\alpha\beta}$ is the deviation from the SM prediction, 
dominated for $\sqrt s\ll \Lambda_{\alpha\beta}$ by the interference term: 
\begin{equation}
\Delta\sigma_{\alpha\beta}\equiv
\sigma_{\alpha\beta}-\sigma_{\alpha\beta}^{\rm SM}\simeq
2 N_C\, \sigma_{\rm pt}
\left(Q_e\, Q_f+g_{\alpha}^e\, g_{\beta}^f\,\chi_Z\right)
\frac{s\eta_{\alpha\beta}}{\alpha\Lambda_{\alpha\beta}^2},
\label{deltasig}
\end{equation}
and $\delta\sigma_{\alpha\beta}$ is the expected experimental uncertainty on 
$\sigma_{\alpha\beta}$, combining statistical and systematic uncertainties.

For example, adding uncertainties in quadrature, the uncertainty on 
$\sigma_{\rm LL}$, indirectly measured {\it via} $\sigma_1$
and $\sigma_2$ (see Eqs.~(\ref{SLL}) and (\ref{sigmap})), 
is given by
\begin{eqnarray}
\label{uncet}
\left(\delta \sigma_{LL}\right)^2
=a^2(z^*)\left ( \frac{1+P}{P} \right )^2 (\delta\sigma_1 (z^*,-P))^2
+a^2(z^*) \left ( \frac{1-P}{P} \right )^2 (\delta\sigma_1 (z^*,P))^2
\nonumber \\
+b^2(z^*) \left ( \frac{1+P}{P} \right )^2 (\delta\sigma_2 (z^*,-P))^2
+b^2(z^*) \left ( \frac{1-P}{P} \right )^2 (\delta\sigma_2 (z^*,P))^2, 
\end{eqnarray}
where
\begin{equation}
a(z^*)=\frac{8-(1-z^*)^3}{6(1-{z^*}^2)}, \qquad
b(z^*)=-\frac{(1-z^*)^3}{6(1-{z^*}^2)} .
\end{equation}
Analogous expressions hold for the combinations related to the uncertainties
$\delta\sigma_{\rm RR}$, $\delta\sigma_{\rm LR}$ and $\delta\sigma_{\rm RL}$.
Numerically, in the situation of small deviations from the SM we are 
considering, we can use to a very good approximation the SM predictions for 
the cross sections $\sigma_{1,2}$ to assess the expected $\delta\sigma_{1,2}$ 
and therefore of the uncertainties $\delta\sigma_{\alpha\beta}$ in the 
denominator of (\ref{signif}). Basically, the directly measured integrated 
cross sections $\sigma_{1,2}$ of Eqs.~(\ref{sigma1}) and (\ref{sigma2}) and, 
correspondingly, the uncertainties $\delta\sigma_{\alpha\beta}$, are dependent
on the value of $z^*$, which can be considered in general as an input 
parameter related to given experimental conditions (see, {\it e.g.}, 
Eq.~(\ref{uncet})). Since the deviation 
$\Delta\sigma_{\alpha\beta}$ of Eq.~(\ref{deltasig}) is independent of $z^*$, 
the full sensitivity of a given helicity cross section to the relevant 
contact-interaction parameter is determined by the corresponding size and 
$z^*$ behavior of the uncertainty $\delta\sigma_{\alpha\beta}$. Then, the 
optimization would be obtained by choosing for $z^*$ the value $z^*_{\rm opt}$
where the uncertainty $\delta\sigma_{\alpha\beta}$ becomes minimum, 
{\it i.e.}, where the corresponding sensitivity Eq.~(\ref{signif}) has a 
maximum. As anticipated, we estimate the required $z^*$ behavior from the 
known SM cross sections. 

Combining, again in quadrature, statistical and
systematic uncertainties on $\sigma_{1,2}$, we have:
\begin{equation}
\label{delsi1}
(\delta\sigma_i)^2\simeq(\delta\sigma_i^{\rm SM})^2=
\frac{\sigma_i^{\rm SM}}{\epsilon\, \Lumint}
+\left(\delta^{\rm sys}\sigma_i^{\rm SM}\right)^2.
\end{equation}

Numerically, for $\sigma_{1,2}$ we take into account the expected 
identification efficiencies, $\epsilon$ \cite{Damerell} and the systematic 
uncertainties, $\delta^{\rm sys}$, on the various fermionic final states, for 
which we assume: for leptons: $\epsilon=95\%$ and $\delta^{\rm sys}=0.5\%$;
for $b$ quarks: $\epsilon=60\%$ and $\delta^{\rm sys}=1\%$;
for $c$ quarks: $\epsilon=35\%$ and $\delta^{\rm sys}=1.5\%$. As
concerns the systematic uncertainty, we assume the same $\delta^{\rm sys}$ 
for $i=1,2$, and independent of $z^*$ in the relevant angular range.   

We consider the LC with the following options: $\sqrt{s}=0.5$~TeV,
$\Lumint=50\ \mbox{fb}^{-1}$, $P=0.8$ and $|\cos\theta|\le 0.99$.  
We assume half the total integrated luminosity quoted above for both 
values of the electron polarization, $P_e=\pm P$.
As an example, the relative uncertainties, 
$\delta\sigma_{\alpha\beta}/\sigma_{\alpha\beta}$,
on the helicity cross sections for the process $e^+e^-\to\mu^+\mu^-$, are 
shown as functions of $z^*$
in Fig.~1. The optimal kinematical parameters  $z^*_{\rm opt}$ where the 
sensitivity of leptonic process is a maximum, can easily be obtained from 
this figure. The corresponding dependences for quark final states are 
analogous.

\begin{figure}[t]
\centerline
{\epsfig{figure=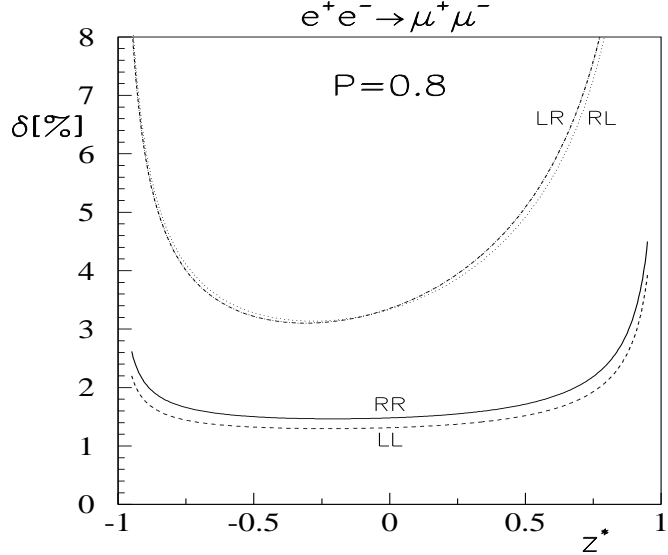,height=8cm,width=10cm}}
\caption{
The uncertainty on the helicity cross sections $\sigma_{\alpha\beta}$ 
in the SM
as a function of $z^*$ for the process $e^+e^-\to\mu^+\mu^-$ at 
$\sqrt{s}=0.5$~TeV, $\Lumint=50\ \mbox{fb}^{-1}$, $P=0.8$, 
$\epsilon=95\%$ and
$\delta^{\rm sys}=0.5\%$. Radiative corrections are included.}
\end{figure}

In order to assess the increase of sensitivity obtained by optimization,
one should compare the corresponding uncertainties at $z^*_{\rm opt}$ with 
those obtained without optimization, at $z^*_\pm$ of Eq.~(\ref{zpm}). 
Fig.~1, show that, in the LL and RR cases, optimization results in a
rather modest increase of sensitivity and of the corresponding discovery 
limits on $\Lambda_{\rm RR}$ and $\Lambda_{\rm LL}$ (by a few percent), 
since the $z^*$ behavior of the uncertainty is rather flat. 
Conversely, in the LR and RL cases optimization can substantially increase 
the sensitivity and the corresponding reachable lower bounds on 
$\Lambda_{\rm LR}$ and $\Lambda_{\rm RL}$ (up to a factor of about 2 
for the $c\bar c$ case).

The bounds on the contact interaction parameters can be obtained by 
using $\chi^2$ procedure.
In the numerical analysis presented below, we take three different values 
of the polarization, $P=$ 1, 0.8, 0.5, in order to study this dependence.
This is a reasonable variation around the value $P=0.8$ expected at the
LC \cite{Accomando}.
\begin{table}[ht]
\centering
\caption{Contact-interaction reach (in TeV) at
an $e^+e^-$ linear collider with $E_{\rm c.m.}=0.5$~TeV
and $\Lumint=50\,\mbox{fb}^{-1}$, at 95\% C.L.
Radiative corrections are included,
with a cut on the energy of photons emitted in the initial state.
The arrows indicate the increase of sensitivity of the observables
caused by the optimization.}
\medskip
\begin{tabular}{|c|c|c|c|c|c|}
\hline
&&&&& \\
process & $P$ & $\Lambda_{\rm LL}$ & $\Lambda_{\rm RR}$ &
$\Lambda_{\rm LR}$ & $\Lambda_{\rm RL}$ \\
&&&&& \\
\hline
\cline{2-6}
 & 1.0 &$40\rightarrow 41$&$39\rightarrow 40$&
$26\rightarrow 40$&$28\rightarrow 41$ \\
\cline{2-6}
$\mu^+\mu^-$ & 0.8 &$37\rightarrow 38$&$37\rightarrow 38$&
$25\rightarrow 37$&$26\rightarrow  37$
\\ \cline{2-6}
 & 0.5 &$32\rightarrow 32$&$31\rightarrow 32$&
$21\rightarrow 30$&$21\rightarrow 30$
\\ \hline
\hline
 &1.0&$41\rightarrow 42$&$45\rightarrow 47$&
$17\rightarrow 31$&$34\rightarrow 42$
\\ \cline{2-6}
${\overline{b}}b$ &0.8&$40\rightarrow 41$&$38\rightarrow 39$&
$17\rightarrow 29$&$29\rightarrow 38$
\\ \cline{2-6}
&0.5&$36\rightarrow 37$&$29\rightarrow 29$&
$13\rightarrow 25$&$22\rightarrow 31$
\\ \hline
\hline
 &1.0&$32\rightarrow 33$&$36\rightarrow 37$&
$21\rightarrow 32$&$20\rightarrow 30$
\\ \cline{2-6}
${\overline{c}}c$ &0.8&$31\rightarrow 32$&$32\rightarrow 33 $&
$20\rightarrow 31$&$18\rightarrow 27$
\\ \cline{2-6}
 &0.5&$27\rightarrow 28$&$26\rightarrow 27$&
$18\rightarrow 27$&$15\rightarrow 22$
\\ \hline
\end{tabular}
\label{tab:tab1}
\end{table}
Table~1 shows that the helicity cross sections $\sigma_{\alpha\beta}$
are quite sensitive to contact interactions,
with discovery limits ranging from 40 to 80
times the CM energy at the degree of electron
polarization $P=0.8$.
The best sensitivity occurs for the
$\bar{b}b$ final state, while the worst one is for $\bar{c}c$.
Decreasing the electron polarization from $P=1$ to $P=0.5$ results in
a worsening of the sensitivity by $20-40\%$, depending on the final state. 
Regarding the role of the assumed uncertainties on the
observables under consideration, in the cases of $\Lambda_{\rm LR}$ and
$\Lambda_{\rm RL}$ the expected statistics are such that the uncertainty
turns out to be dominated by the statistical one, and the results have little
sensitivity to the value of the systematic uncertainty. Conversely, in the
cases of $\Lambda_{\rm LL}$ and $\Lambda_{\rm RR}$ the results depend more
sensitively on the assumed value of the systematic uncertainty.
Moreover, as is evident from Eqs.~(\ref{s+}) and (\ref{s-}), a further
improvement in the sensitivity to the various $\Lambda$-scales in Table~1 
would be obtained if both $e^-$ and $e^+$ longitudinal
polarizations were available.
\section{Constraints and resolving power on $Z^\prime$}
As mentioned in the Introduction, also the $s$-channel exchange of a new,
very massive, neutral gauge boson $Z'$ with $\sqrt{s}\ll M_{Z'}$
can be identified to a contact interaction Lagrangian of the kind
in Eq.~(\ref{lagra}). Specifically, the contribution of
the $Z'$-mediated helicity amplitudes, to be added to the SM ones,
takes the form:
\begin{equation}
A_{\alpha\beta}(Z^\prime)=
{g^{\prime}}^e_{\alpha}\,{g^{\prime}}^f_{\beta}\,\chi_{Z^\prime}
\simeq
-{g^{\prime}}^e_{\alpha}\,{g^{\prime}}^f_{\beta}\,\frac{s}{M_{Z^\prime}^2}
\left(1+\frac{s}{M_{Z^\prime}^2}+...\right),
\label{amplitz}
\end{equation}
where $\chi_{Z^\prime}=s/(s-M^2_{Z^\prime})$ is the $Z^\prime$ propagator,
and ${g^{\prime}}^e_{\alpha}$, ${g^{\prime}}^f_{\alpha}$
are the fermionic couplings of the $Z^\prime$ \cite{Osland,Babich}:
\begin{equation}
{g^\prime}^f_L=\frac{g_{Z^\prime}}{e}\,L_{Z^\prime}^f, \qquad
{g^\prime}^f_R=\frac{g_{Z^\prime}}{e}\,R_{Z^\prime}^f.
\label{left}
\end{equation}
Comparing  Eqs.~(\ref{amplit}) and (\ref{amplitz}), one finds
\begin{equation}
\label{contz}
\frac{\eta_{\alpha\beta}}{\Lambda_{\alpha\beta}^2}
\approx
-g^{\prime e}_{\alpha}\,g^{\prime f}_{\beta}\,\frac{\alpha}{M_{Z^\prime}^2}.
\end{equation}
Thus, in the case of no obrserved signal, i.e., no deviation of 
$\sigma_{\alpha\beta}$ from the SM prediction within the experimental accuracy,
at given $M_{Z^\prime}$
one can directly obtain model-independent bounds on the 
fermionic chiral couplings to the $Z^\prime$ from 
Eqs.(\ref{amplitz})-(\ref{contz}) and the constraints listed in Table~1.

If a $Z^\prime$ is indeed discovered, perhaps at a hadron machine, it 
becomes interesting to measure as accurately as possible its couplings 
and mass at the LC, and make tests of the various extended gauge models. 
Another interesting question is the potential of the leptonic process 
(\ref{proc}) to identify the $Z^\prime$ model underlying the measured 
signal, through the measurement of the helicity cross sections, e.g. 
$\sigma_{RR}$ and $\sigma_{LL}$. Such cross sections only depend on 
the relevant leptonic chiral coupling and on $M_{Z^\prime}$, so that such 
resolving power clearly depends on the actual value of the $Z^\prime$ mass. 
In Figs.~2a and 2b we show 
this dependence for the $E_6$ and the $LR$ models of interest here. In these 
figures, the horizontal lines represent the values of the couplings predicted
by the various models, and the lines joining the upper and the lower ends of 
the vertical bars represent the expected experimental uncertainty at the 
95\% CL. The intersection of the lower such lines with the $M_{Z^\prime}$ 
axis determines the discovery reach for the corresponding model: larger 
values of $M_{Z^\prime}$ would determine a $Z^\prime$ signal smaller than 
the experimental uncertainty and, consequently, statistically invisible. 
Also, Figs.~2a and 2b show the complementary roles of $\sigma_{LL}$ and 
$\sigma_{RR}$ to set discovery limits: while $\sigma_{LL}$ is mostly 
sensitive to the $Z^\prime_\chi$ and has the smallest sensitivity to the 
$Z^\prime_\eta$, $\sigma_{RR}$  provides the best limit for the
$Z^\prime_{LR}$ and the worst one for the $Z^\prime_\chi$. 

\begin{figure}[t]
\centerline
{\epsfig{figure=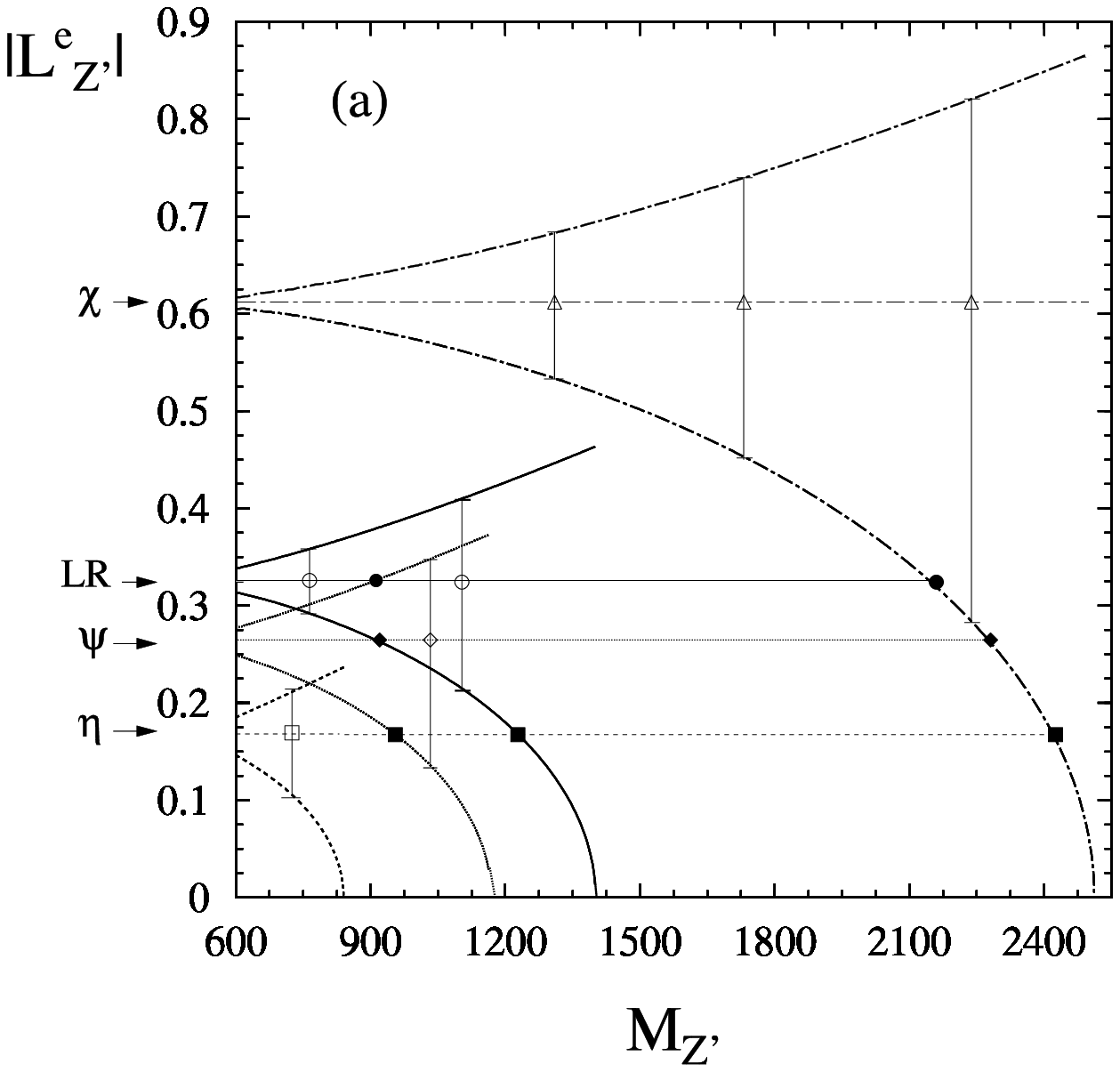,height=8cm,width=10cm}}
\centerline
{\epsfig{figure=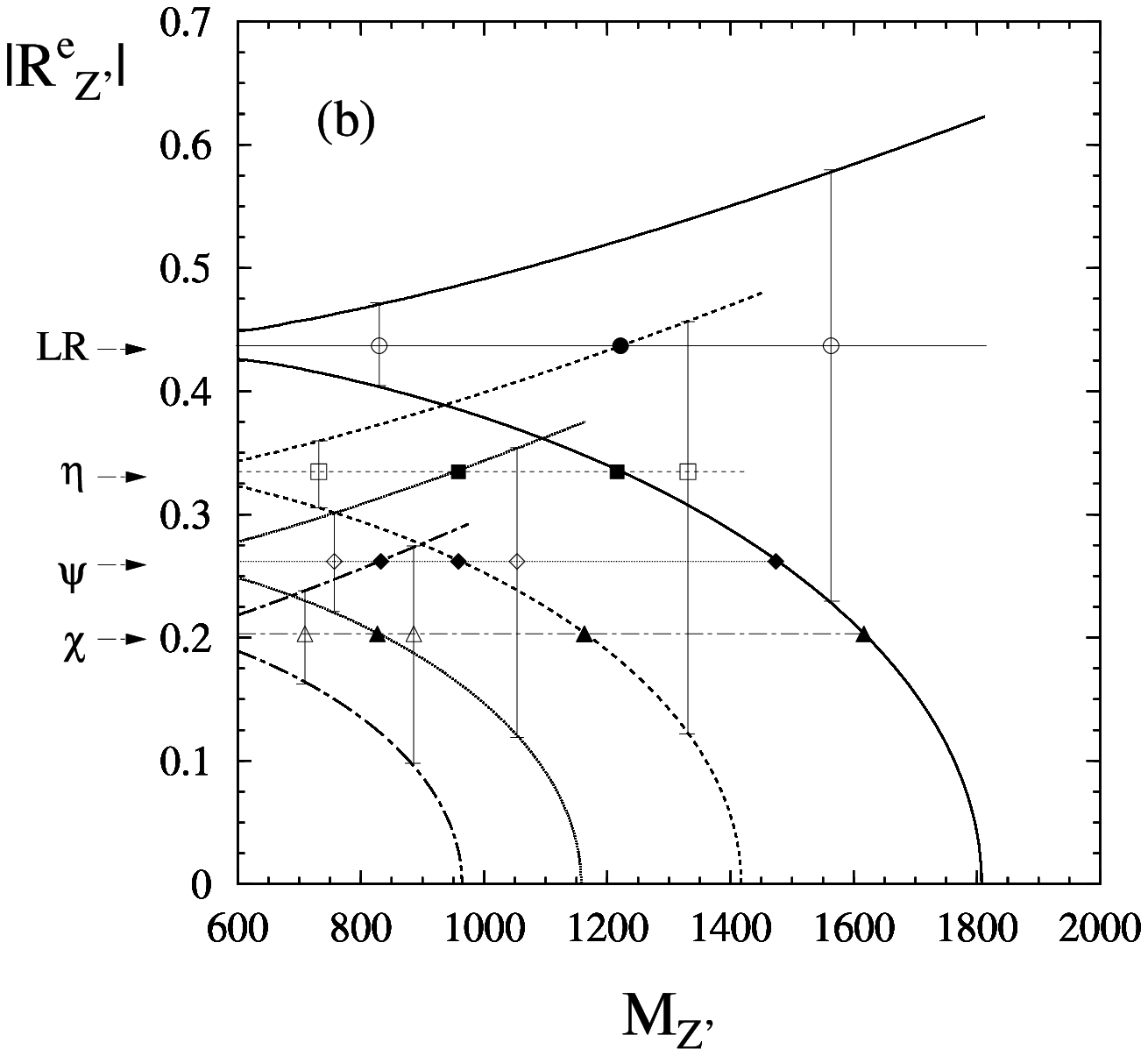,height=8cm,width=10cm}}
\caption{
Resolution power at 95\% C.L. for the absolute value of the 
leptonic $Z^\prime$ couplings, $\vert L^e_{Z^\prime}\vert$ (a) and 
$\vert R^e_{Z^\prime}\vert$ (b), 
as a function of $M_{Z^\prime}$, obtained 
from $\sigma_{LL}$ and $\sigma_{RR}$, respectively, 
in process $e^+e^-\to l^+l^-$.  The error bars 
combine statistical and systematic uncertainties. 
Horizontal lines correspond 
to the values predicted by typical models. 
}
\end{figure}

As Figs.~2a and 2b show, the different models can be distinguished by means
of $\sigma_\pm$ as long as the uncertainty of the coupling of one model does 
not overlap with the value predicted by the other model. Thus, the
identification power of the leptonic process (\ref{proc}) is determined by
the minimum $M_{Z^\prime}$ value at which such `confusion region' starts. 
For example, Fig.~2a shows that the $\chi$ model cannot be distinguished from the LR, 
$\psi$ and $\eta$ models at $Z^\prime$ masses larger than 2165 GeV, 2270 GeV 
and 2420 GeV, respectively. The identification power for the typical models 
are indicated in Figs.~2a and 2b by the symbols circle, diamond, square and 
triangle.

In the case of process (\ref{proc}) with ${\bar q}q$ pair production (with 
$q=c,\, b$), the analysis is complicated by the fact that the relevant 
helicity amplitudes depend on three parameters ($g^{\prime e}_\alpha$,
$g^{\prime q}_\beta$ and $M_{Z^\prime}$) instead of two. Nevertheless, there 
is still some possibility to derive general information on the $Z^\prime$ 
chiral couplings to quarks.
As an illustrative example, in Fig.~3 we depict 
the bounds from the process $e^+e^-\to{\bar b}b$ in the 
($L^e_{Z^\prime}$,$L^b_{Z^\prime}$) and ($L^e_{Z^\prime}$,$R^b_{Z^\prime}$) 
planes for the $Z^\prime$ of the $\chi$ model, with 
$M_{Z^\prime}= 1\,{\rm TeV}$. Taking into account two-fold 
ambiguity, the allowed regions are the ones included within the two sets of 
hyperbolic contours in the upper-left and in the lower-right corners of 
Fig.~3. Then, to get finite regions for the quark couplings, one must 
combine the hyperbolic regions so obtained with the determinations of the 
leptonic $Z^\prime$ couplings from the leptonic process (\ref{proc}), 
represented by the two vertical strips. The corresponding shaded areas 
represent the determinations of $L^b_{Z^\prime}$, while the hatched 
areas are the determinations of $R^b_{Z^\prime}$. Notice that, in general, 
there is the alternative possibility of deriving constraints on quark
couplings also in the case of right-handed electrons, namely, 
from the determinations of
the pairs of couplings ($R^e_{Z^\prime}$,$L^b_{Z^\prime}$) and
($R^e_{Z^\prime}$,$R^b_{Z^\prime}$). However, as observed with regard to the
previous analysis of the leptonic process, the sensitivity to the 
right-handed electron coupling turns out to be smaller than for 
$L^e_{Z^\prime}$, so that the corresponding constraints are weaker. 

\begin{figure}[t]
\centerline
{\epsfig{figure=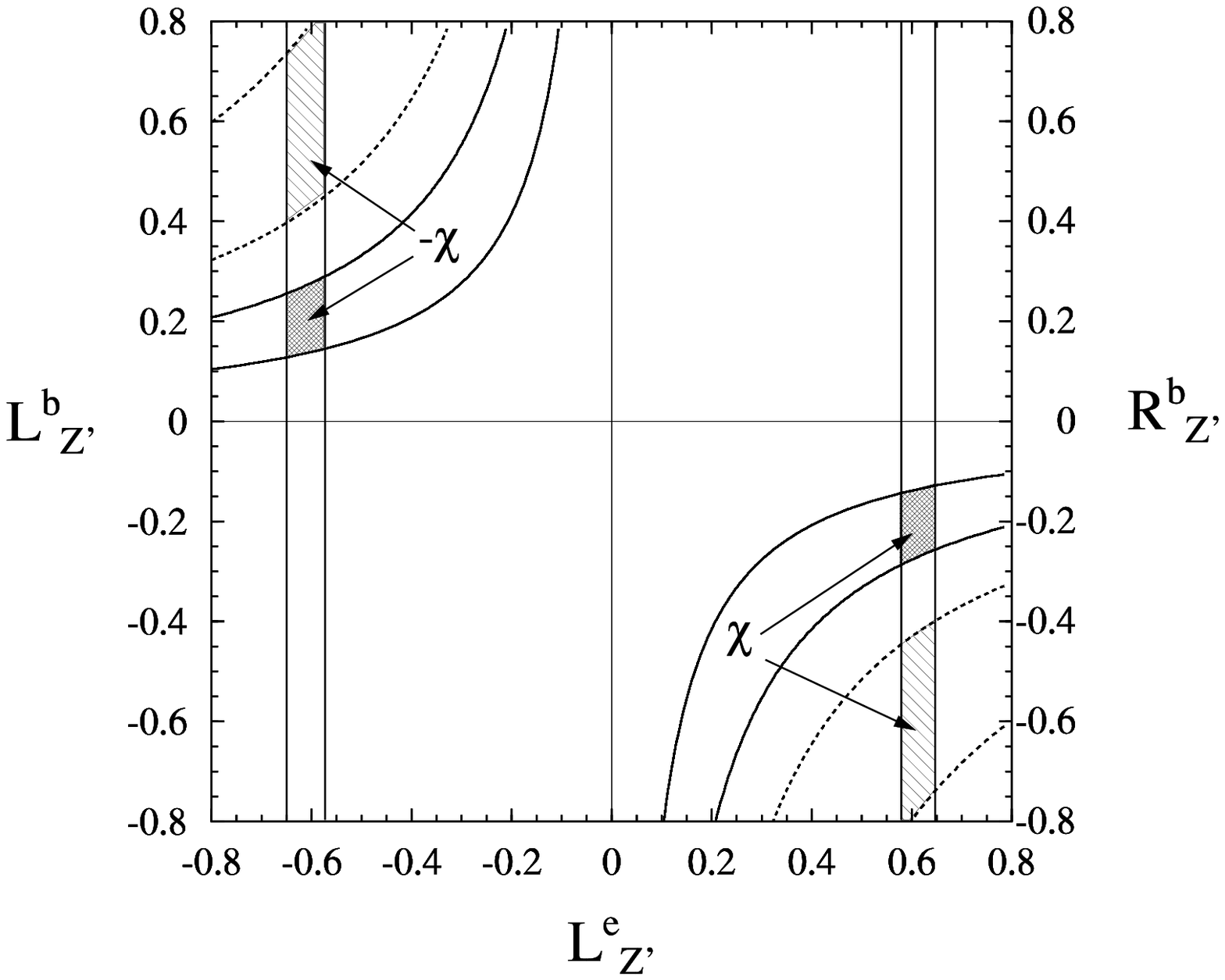,height=8cm,width=10cm}}
\caption{
Allowed bounds at 95\% C.L. on $Z^\prime$ couplings with $M_{Z^\prime}=1$ TeV
($\chi$ model) in the two-dimension planes
($L^e_{Z^\prime}$,$L^b_{Z^\prime}$) and ($L^e_{Z^\prime}$,$R^b_{Z^\prime}$)
obtained from helicity cross sections $\sigma_{LL}$ (solid lines) and
$\sigma_{LR}$ (dashed lines), respectively.
The shaded and hatched regions are derived from the combination of
$e^+e^-\to l^+l^-$ and $e^+e^-\to\bar{b}b$ processes.
}
\end{figure}
\section{Summary}
We emphasize that the measurement of the helicity cross sections 
of the process (\ref{proc}) with
optimal kinematical cuts could substantially increase their
sensitivity to contact interaction parameters and could give crucial, 
model-independent information on the chiral structure of such new
interactions. 

As an application of the proposed approach at LC, we 
study the sensitivity to $Z^\prime$.     
In the case of no observed signal, 
one can directly obtain model-independent bounds on the leptonic chiral 
couplings of the $Z^\prime$ from $e^+e^-\to l^+l^-$ and on the products of 
lepton-quark chiral couplings from $e^+e^-\to {\bar q}q$ (with 
$l=\mu,\tau$ and $q=c,b$). 
In the case $Z^\prime$ manifestations are observed as deviations 
from the SM, the role of 
$\sigma_{\alpha\beta}$ is more interesting, specially as 
regards the problem of identifying the various models as potential 
sources of such non-standard effects. 
Indeed, in principle, they provide a unique  possibility to 
disentangle and extract numerical values for the chiral couplings of 
the $Z^\prime$ in a general way, avoiding the danger of cancellations, 
so that $Z^\prime$ model predictions can be tested. 

\section*{Acknowledgements}
I would like to thank P. Osland, N. Paver and A.A. Babich for the fruitful 
and enjoyable collaboration on the subject matter covered here.

\end{document}